\documentclass{emulateapj}
\usepackage{epsfig}
\usepackage{amsmath}
\usepackage{amssymb}
\usepackage{natbib}
\usepackage{setspace}
\usepackage{graphicx}

\begin{document}
\title{Reconstructing the stellar mass distributions of galaxies using S$^4$G~IRAC~3.6~and~4.5~$\mu \textsc{m}$ images: II. the conversion from light to mass}
\author{Sharon E. Meidt\altaffilmark{1}, Eva Schinnerer\altaffilmark{1}, Glenn van de Ven\altaffilmark{1}, Dennis Zaritsky\altaffilmark{2}, Reynier Peletier\altaffilmark{3}, Johan Knapen\altaffilmark{4,5}, 
Kartik Sheth\altaffilmark{6,7,8}, Michael Regan\altaffilmark{9}, Miguel Querejeta\altaffilmark{1}, Juan-Carlos Mu\~noz-Mateos\altaffilmark{10}, Taehyun Kim\altaffilmark{6}, Joannah L. Hinz\altaffilmark{2}, 
Armando Gil de Paz\altaffilmark{11}, E. Athanassoula\altaffilmark{12}, Albert Bosma\altaffilmark{12}, Ronald J. Buta\altaffilmark{13}, Mauricio Cisternas\altaffilmark{4,5}, Luis C. Ho\altaffilmark{14}, Benne Holwerda\altaffilmark{15}, Ramin Skibba\altaffilmark{2}, E. Laurikainen$^{16,17}$,
H. Salo$^{16}$, D. A. Gadotti$^{10}$, Jarkko Laine\altaffilmark{16,17}, S. Erroz-Ferrer$^{4, 5}$, S\'{e}bastien Comer\'{o}n\altaffilmark{16,17}, K. Men\'{e}ndez--Delmestre$^{18}$, M. Seibert$^{13}$, T. Mizusawa$^{6, 19}$
}

\altaffiltext{1}{Max-Planck-Institut f\"ur Astronomie / K\"{o}nigstuhl 17 D-69117 Heidelberg, Germany}
\altaffiltext{2}{University of Arizona}
\altaffiltext{3}{Kapteyn Astronomical Institute, University of Groningen}
\altaffiltext{4}{Instituto de Astrofisica de Canarias, Spain}
\altaffiltext{5}{Departamento de Astrof\'\i sica, Universidad de La Laguna, Spain}
\altaffiltext{6}{National Radio Astronomy Observatory}
\altaffiltext{7}{Spitzer Science Center}
\altaffiltext{8}{California Institute of Technology}
\altaffiltext{9}{Space Telescope Science Institute}
\altaffiltext{10}{European Southern Observatory, Chile}
\altaffiltext{11}{Universidad Complutense Madrid}
\altaffiltext{12}{Laboratoire d'Astrophysique de Marseille (LAM)}
\altaffiltext{13}{University of Alabama}
\altaffiltext{14}{The Observatories of the Carnegie Institution for Science}
\altaffiltext{15}{European Space Agency Research Fellow (ESTEC)}
\altaffiltext{16}{University of Oulu, Finland}
\altaffiltext{17}{Finnish Centre for Astronomy with ESO (FINCA), University of Turku}
\altaffiltext{18}{Observat{\'o}rio do Valongo, Universidade Federal do Rio de Janeiro}
\altaffiltext{19}{Florida Institute of Technology}

\date{\today}
\begin{abstract}
We present a new approach for estimating the 3.6 $\mu m$ stellar mass-to-light ratio $\Upsilon_{3.6}$ in terms of the [3.6]-[4.5] colors of old stellar populations.  Our approach avoids several of the largest sources of uncertainty in existing techniques using population synthesis models.  By focusing on mid-IR wavelengths, we gain a virtually dust extinction-free tracer of the old stars,   avoiding the need to adopt a dust model to correctly interpret optical or optical/NIR colors normally leveraged to assign the mass-to-light ratio $\Upsilon$.  By calibrating a new relation between NIR and mid-IR colors of giant stars observed in GLIMPSE we also avoid the discrepancies in model predictions for the [3.6]-[4.5] colors of old stellar populations due to uncertainties in the molecular line opacities assumed in template spectra.  
We find that 
the [3.6]-[4.5] color, which is driven primarily by metallicity, provides a tight constraint on $\Upsilon_{3.6}$, which varies intrinsically less than at optical wavelengths.  
The uncertainty on $\Upsilon_{3.6}$ of $\sim$0.07 dex due to unconstrained age variations    
marks a significant improvement on existing techniques for estimating the stellar M/L with shorter wavelength data.  A single $\Upsilon_{3.6}$=0.6 (assuming a Chabrier IMF), independent of [3.6]-[4.5] color, is also feasible as it can be applied simultaneously to old, metal-rich and young, metal-poor populations, and still with comparable (or better) accuracy ($\sim$0.1 dex) as alternatives.    
We expect our $\Upsilon_{3.6}$ to be optimal for mapping the stellar mass distributions in S$^4$G galaxies, for which we have developed an Independent Component Analysis technique to first isolate the old stellar light at 3.6 $\mu m$ from non-stellar emission (e.g. hot dust and the 3.3 PAH feature).  Our estimate can also be used to determine the fractional contribution of non-stellar emission to global (rest-frame) 3.6 $\mu m$ fluxes, e.g. in WISE imaging, and establishes a reliable basis for exploring variations in the stellar IMF.    

\end{abstract}
\section{Introduction\label{sec:intro}}
Accurate maps of the stellar mass distribution in nearby galaxies are essential for charting the dynamical influence of stellar structures over time.  They provide a key perspective on the evolution in the amount and distribution of baryons from high redshift to the local universe, both as the result of internal, secular processes (e.g. Elmegreen et al. 2007; Zhang \& Buta 2007; Haan et al 2009; Foyle et al. 2010), and given `bottom up' assembly and external (heirarchical accretion) events within the context of CDM cosmology (
e.g., Balcells et al. 2003 and Courteau et al. 1996;  Florido et al. 2001; and see Navarro \& White 1994; Somerville 2002; Governato et al. 2007; Guo \& White 2008).\\
\indent NIR bands are often though to be optimal windows on the old stars that dominate the baryonic mass in galaxies, with images at these wavelengths serving as good relative proxies for stellar mass (Elmegreen \& Elmegreen 1984; Rix \& Zaritsky 1995; Grosbol 1993).  
Yet even in these bands the light is affected by the presence of dust extinction and young stars (albeit less than in the optical), requiring the application of a M/L ratio that incorporates dependencies on metallicity, star formation history, and especially the age of the stellar population (Rix \& Rieke 1993; Rhoads 1998).  
Such estimates have emerged only relatively recently, following the demonstrated correlation between optical colors and the mass-to-light ratio in stellar population synthesis models \citep{bdJ01}.  The state-of-the art, which uses more than one color and incorporates a model of the dust, is optimal for use in 2D and has greatly reduced uncertainties \citep{z09}.  But, at present few such maps exist.  

\indent We have begun an effort to map the stellar mass distribution in nearby galaxies in the Spitzer Survey of Stellar Structure in Nearby Galaxies (S$^4$G), an un-paralleled inventory of stellar mass and structure in $\sim$2300 nearby galaxies ($D$$<$40 Mpc; Sheth et al. 2010).  The survey consists of deep imaging 
at 3.6 $\mu m$ and 4.5 $\mu m$ where the light traces the old stars, and where the effects of 
dust extinction and star formation activity (e.g. young stars, HII regions; see \citealt{sheth} and references therein) are at a minimum compared to ground-based images in either optical or NIR wavebands. 

Our method begins by using Independent Component Analysis (ICA) to isolate the old stellar light (from K and M giants) from 
the `contaminating' emission from several other sources: the 3.3 $\mu m$ PAH emission feature in channel 1, accessory non-stellar continuum emission (i.e. from hot dust and PAH), and lower-M/L asymptotic giant branch (AGB) and red supergiant (RSG) stars.  
These three main contaminants can lead to an over-estimation of stellar mass when using uncorrected 3.6 $\mu m$ and/or 4.5 $\mu m$ images (\citealt{meidtpaperI}; hereafter Paper I).  Corrections for this `contaminating' emission are especially critical for mass estimates, since this emission is not incorporated into SED libraries at the core of now-standard techniques for estimating the stellar M/L (e.g. \citealt{bdJ01}; \citealt{z09}).   

Once contaminants are removed with ICA we obtain maps of the underlying distribution of old stellar light in S$^4$G images, with [3.6]-[4.5] colors that are consistent with those of M and K giants (\citealt{pahre}; \citealt{hunter}).   To generate 2D maps of the stellar mass distribution in principle thus requires applying a M/L that is appropriate for an old, dust-free population of stars.  

Tight constraints on the stellar M/L at 3.6$\mu m$ are possible using stellar population synthesis (SPS) models to link, e.g., optical and/or NIR colors to metallicity and age (or star formation history; SFH) variations in the underlying stellar population.  Paired with this information, contaminant-corrected S$^4$G images should be readily convertible into high-quality mass maps.  
But even in the latest generation of models there are lingering uncertainties:  1) convergence on the TP-AGB phase of stellar evolution that dominates NIR light at $\sim$1 Gyr has yet to be reached among different models (cf. Charlot \& Bruzual 2007, \citealt{maraston05} and \citealt{lancon}), 
2) differences in the molecular line opacities assumed in template spectra lead to a variety of model predictions at mid-IR wavelengths (e.g. Peletier et al. 2012), and 3) realistic models of dust extinction (as well as the reflection and scattering of stellar light by dust) are difficult to implement at the pixel-by-pixel level, given the variety of dust geometries \citep{z09}.   

In this paper we explore estimating the M/L at 3.6$\mu m$ in a way that is least susceptible to the uncertainties that plague state-of-the-art mass estimation techniques.  The key is our focus on mid-IR wavelengths where the emission is dominated by the oldest stars and dust extinction is minimal.  

In $\S$ \ref{sec:ICA} we describe our ICA technique to isolate the old stellar light from non-stellar emission at 3.6 $\mu m$.  Then in $\S$ \ref{sec:MLdescription} we demonstrate that the variation in stellar $\Upsilon_{3.6}$ is intrinsically quite low and, in this case, we argue that the IRAC [3.6]-[4.5] colors of the old stars provide sufficient constraint on the M/L ($\S$ \ref{sec:withcolor}).  To demonstrate this we use SPS models, setting out first to improve their predictive power at mid-IR wavelengths.  Using a new calibration of the relation between the NIR and mid-IR colors of GLIMPSE giants (Appendix A), we link SPS models to realistic [3.6]-[4.5] colors in $\S$ \ref{sec:iraccolor}.  This allows us to explore the dependence of age/SFH (including bursts) and metallicity on the relation between $\Upsilon_{3.6}$ and [3.6]-[4.5] color, which we present in $\S$ \ref{sec:therelation}. 

Our approach for constructing 2D mass maps, namely, using information exclusively from 3.6 and 4.5 $\mu m$ images revealed by ICA, is an approach that can be applied to the greatest number of galaxies in the S$^4$G sample without relying on ancillary optical and/or NIR data. 

\section{ICA correction of S$^4$G images}
\label{sec:ICA}
Mass maps for much of the S$^4$G sample will necessarily rely on information from the two lowest IRAC wavebands, the only bands available during the Spitzer warm mission.  Given this constraint, we choose to remove contaminants from these images using an Independent Component Analysis (ICA) technique that requires only the information available in these two bands and solves for both the fluxes of, and the scaling between (i.e. color), each of two independent sources.  

\subsection{Applying ICA\label{sec:sources}}
As described in Paper I, we assume that emission in each pixel mapped at each of two frequencies $\nu_i$ is a linear combination of the emission from $N=2$ different sources (stellar and non-stellar) such that, for $P$ samples (viz. image pixels) in $M$=2 frequency channels,
\begin{equation}
\mathbf{x}=\mathbf{A}\cdot\mathbf{s}\label{eq:ica}
\end{equation}
where $\textbf{x}$ is the $M$$\times$$P$ measurement set, $\textbf{A}$ is an $M$$\times$$N$ mixing matrix containing the coefficients of the combination and the $N$$\times$$P$ matrix $\textbf{s}$ represents the source signals, under the assumption that the sources $\textbf{S}$ are separable functions of location and frequency (i.e. $s_j$=$S_j(x,y)$ and $A_{i,j}\propto S_{j}(\nu)$). 

Given our two input channel 1 and 2 images $x_i$, we can apply ICA to extract measurements of at most two signals $s_j$ (each representing either contaminants or emission from the true old stellar disk) in each of $P$ pixels fwhile simultaneously solving for the separation coefficients $A^{-1}_{ij}$.  

Like Principal Component Analysis (PCA), ICA extracts solutions via transformation of the input data, but the sources are statistically independent, rather than orthogonal (with zero covariance).  This grants greater uniqueness to ICA solutions, promising improved descriptions of sources that are the result of distinct physical processes.   

Perhaps the strongest advantage of ICA is that it allows us to retain and exploit information about the different spectral shapes of the stars and dust between 3.6 and 4.5 $\mu  m$, rather than being forced to make a priori assumptions about the properties of either the stars or the contaminating emission (e.g., imposing a particular set of separation coefficients $A^{-1}_{ij}$ given some empirical recipe).  This is a particularly favorable alternative to 1) subtracting scaled versions of the 8$\mu m$ image (which requires prior knowledge of the relation between the 3.3$\mu m$ PAH emission feature and the 8$\mu m$ -- and how, in particular, this varies depending on the size and ionization state of the dust) or 2) a simple scaling and subtraction of the 3.6 and 4.5 $\mu  m$ images.  
Given the difference in SED shapes for the old stars and dust, there is not a single scaling between the emission at 3.6 $\mu m$ and at 4.5 $\mu m$ that can be applied uniformly to all galaxies throughout the Hubble sequence.  Combining the two images would furthermore require the loss of the [3.6]-[4.5] color, a key diagnostic of the properties of the old stellar population.  By avoiding such measures, we gain a unique tracer of the non-stellar emission (and the thermal and non-thermal sources of dust heating; paper I) as well as a measure of the color of the old stars, which we leverage in this paper to assign the M/L$_{3.6}$.  

In addition, ICA in principle allows much more information in individual structural components, like spiral arms, to be maintained in contaminant-cleaned maps than is supplied by other commonly-employed mass estimation methods.  
Alternative schemes for the identification and/or correction of these features in IRAC images (Kendall et al. 2008; \citealt{hunter}; Bolatto et al. 2007; de Blok et al. 2008) either require the loss of 2D information, or employ auxiliary data (e.g. 8$\mu m$, UV, optical images) that is not uniformly available for the full S$^4$G sample.  ICA uses only the information from the first two IRAC bands and thus provides a correction for non-stellar emission that can be made uniformly throughout the sample.

\subsection{Sources identified with ICA\label{sec:sources}}
\indent 
As outlined in Paper I (and described in detail by Querejeta et al. 2014), our application of ICA with additional post-processing yields two maps: a corrected map of the old stellar light (a clean, smoothed version of the uncorrected image) and a map of significant contaminating non-stellar emission (containing most, if not all, of the bright and knotty features tracing, e.g., star formation in spiral arms).  

\subsubsection{Non-stellar emission}
By comparing with the IRAC 8$\mu m$ image, which is dominated by PAH emission, in Paper I we demonstrated that the main sources of contamination expected in 3.6 $\mu m$ images are well-detected with ICA.  These contaminants are defined as all sources of emission apart from the oldest 
stars: PAH, hot dust, and intermediate-age RSG and AGB stars.  Whereas old stars typically exhibit blue colors in the range -0.15$<$[3.6]-[4.5]$<$-0.02, the emission from non-stellar sources are characterized by colors [3.6]-[4.5]$\sim$0.3 and redder (Meidt et al 2012; Querejeta et al. 2014).   

The fractional contribution of non-stellar emission to the integrated 3.6 $\mu m$ flux of galaxies in S$^4$G can be significant, as high as 30\% (Meidt et al. 2012; Querejeta et al. 2014).  This emission covers roughly one third of the disk area, although the degree of spatial contamination depends on the type of galaxy and the nature of the non-stellar emission.  Dust emission tracing star formation in low-mass, late-type disks with flocculent spirals, for example, tends to be more diffuse and uniformly distributed.  In contrast, the star formation at the source of dust heating in more massive disks with coherent spiral arms tends to be strongly locally organized and thus typically arises from a smaller fraction of the disk area (Querejeta et al. 2014)

\subsubsection{Old stellar light maps}
By removing non-stellar emission in S$^4$G images with ICA we uncover information related to the old stellar population that was not initially accessible with only the 3.6 and 4.5 $\mu m$ bands, i.e. as a result of non-stellar contamination.   Within contaminated regions ($\sim$30\% of the disk area) we obtain new flux and color information for the underlying old stars.  
In all other initially contaminant-free locations (typically 70\% of the disk) the native flux and color information tracing the properties of the stellar population are retained.  
Only in cases with pervasive non-stellar emission, in which a single [3.6]-[4.5] is assigned to a larger fraction of the disk, do we lose full 2D color information to constrain the properties of the stellar population.  As will be discussed in $\S$, however, this is not especially adverse for stellar mass estimation: changes in the stellar M/L due to stellar age and metallicity variations are minimal, and a single (color-independent) M/L may even be preferred.  

Removing non-stellar emission with ICA introduces only at most 15\% uncertainty (Querejeta et al. 2014), which improves on the uncertainty that would be associated with the presence of that emission (roughly 30\%).  
Images of the underlying old stellar light at 3.6 $\mu m$ better match expectations for an old, dust-free stellar population than those from which the contaminant emission has not been removed (paper I).  
In this case, to convert from light to mass, we only have to apply a {\it stellar} M/L, rather than one that accounts for the presence of dust emission tracing young stars/recent star formation.  
In the upcoming sections, we outline our approach to determine the optimal conversion between the light at 3.6 $\mu m$ and stellar mass.  

\begin{figure*}[t]
\begin{centering}
\begin{tabular}{c}
\includegraphics[width=.95\linewidth]{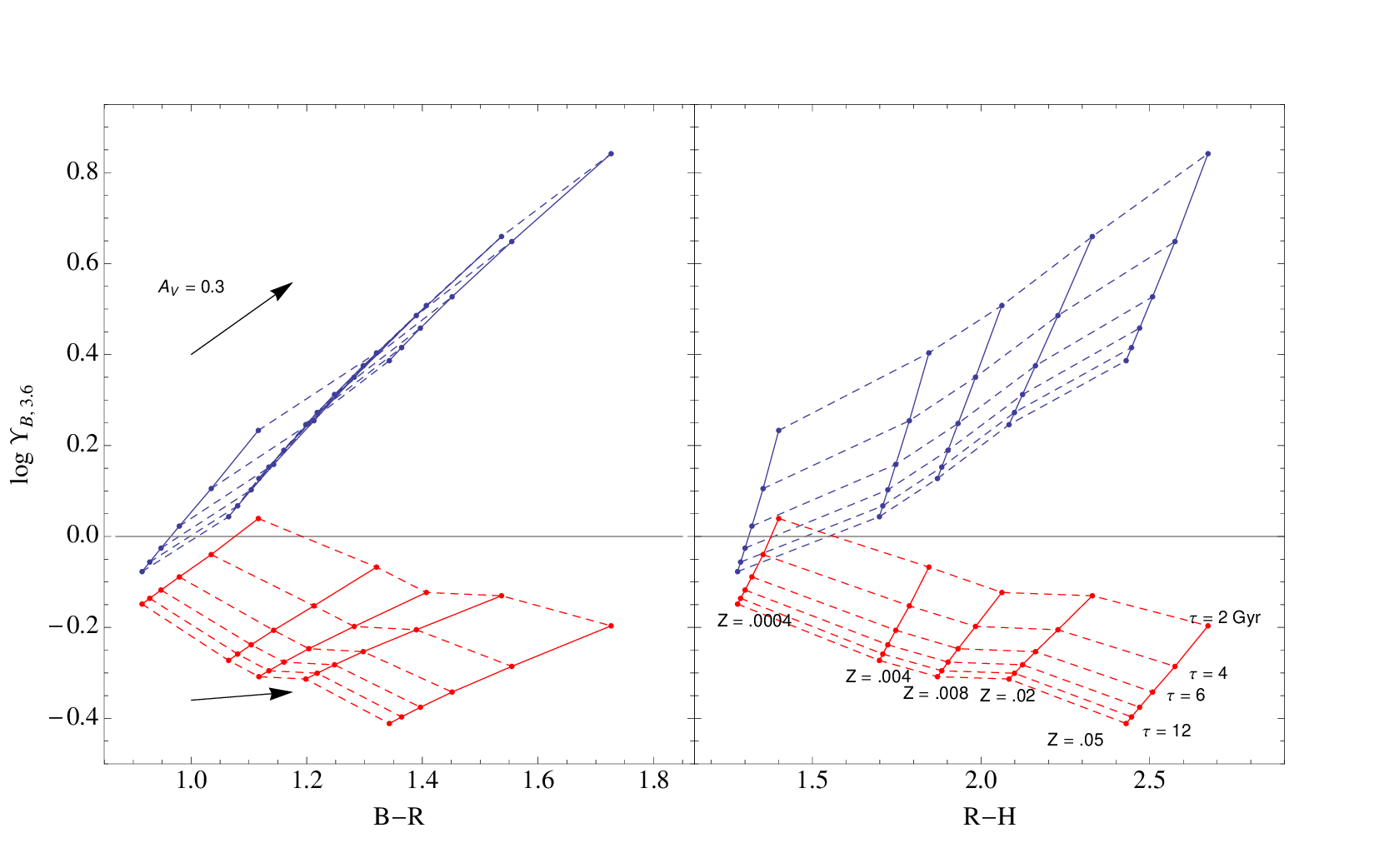}
\end{tabular}
\caption{Trends in stellar M/L ratios (in solar units $M_{\odot}/L_{\odot}$) with optical and optical-NIR color for simple exponential SFH models of age 13 Gyr from BC03.  The blue lines (top grid) show B-band $\Upsilon_B$, and the red lines (bottom grid) show 3.6 $\mu m$ $\Upsilon_{3.6}$.  Models of the same e-folding timescale $\tau$ are connected by dashed lines, while models of the same metallicity $Z$ are connected by solid lines.  Short (long) $\tau$ correspond to older (younger) ages. The black arrows in the top and bottom of the left panel corresponds to V-band extinction A$_V$=0.3, assuming a Galactic extinction law (\citealt{cardelli}; \citealt{indebetouw}).  \label{fig:bothMLgrids}}
\end{centering}
\end{figure*}
\section{General trends in the stellar mass-to-light ratio at 3.6 $\mu m$}\label{sec:MLoverview}
\subsection{Comments on the sources of variation}\label{sec:MLdescription}
As in other NIR bands, the light at 3.6~$\mu m$ is dominated by the old stars that populate the red giant branch (RGB).  This furnishes the powerful advantage that the stellar M/L, which must reflect the mix of these giants stars with the underlying mass-dominant dwarf stars, varies much less with age/SFH and metallicity than at optical wavelengths (Bell \& de Jong 2001).   

Consider that, in a simple stellar population, as stars evolve off the main sequence (and the number of stars left on the main sequence decreases) the RGB is populated by later and later spectral types with decreasing M/L.  The M/L of the stars remaining on the main sequence meanwhile increases, weighted more toward the higher M/L of the late M spectral class.  At the same time, the fraction of giants to dwarfs increases since the average time spent on the RGB is longer than their average main sequence lifetime. These factors keep the NIR M/L for the oldest stars free of the relatively high degree of variation characteristic of optical wavelengths as the population evolves.   
Together with the reduced sensitivity to dust extinction, this is the primary reason these bands are favored as direct stellar mass tracers.    

The photometric properties of stellar populations, which progressively redden as the RGB becomes dominant,\footnote{or more specifically, with time since the start of the last episode of star formation} encode this difference in stellar-M/L with wavelength.  By construction, the mass-to-light ratio at 3.6~$\mu m$ is 
\begin{equation}
\Upsilon_{3.6}=\Upsilon_{\lambda}\frac{10^{-0.4(m^\lambda-m^{3.6})}}{10^{-0.4(M_{\odot}^{\lambda}-M_{\odot}^{3.6})}}
\end{equation}
where, at any wavelength $\lambda$, $\Upsilon_{\lambda}$ is the mass-to-light ratio, $m^{\lambda}$ is apparent magnitude and $M_{\odot}^\lambda$ is the absolute magnitude of the sun.  
The prominence of the RGB that leads to red optical-NIR colors ensures a smaller $\Upsilon_{3.6}$ than in the optical.     
For old stellar populations (redder than the Sun), the variation in $\Upsilon_{3.6}$ compared to $\Upsilon_\lambda$ will also tend to be smaller since 
\begin{equation}
\frac{d\Upsilon_{3.6}}{d\Upsilon_{\lambda}}=\frac{10^{-0.4(m_\lambda-m_{3.6})}}{10^{-0.4(M_{\odot}^{\lambda}-M_{\odot}^{3.6})}}\left(1-0.4\frac{\partial(m_\lambda-m_{3.6})}{\partial\beta}\frac{d\beta}{d\log{\Upsilon_{\lambda}}}\right)
\end{equation}
where variation in parameter $\beta$ (representing, i.e., age or metallicity) is responsible for the change in the properties of the population, both color and mass-to-light ratio.  
Even progressing from the H-band to 3.6 $\mu m$, the variation in the stellar M/L is intrinsically reduced; assuming a color H-[3.6]=0.4 typical of old, dust free population, $\Delta\Upsilon_{3.6}$ will be at most 0.75$\Delta\Upsilon_{H}$.  

In the next section, we explore the variation in mass-to-light ratio in more detail using population synthesis models to follow the evolution in the fraction of giant to dwarf stars, for a given (Chabrier) IMF, as well as the dependence of their combined properties on metallicity and star formation history.   We compare $\Upsilon_{3.6}$ with $\Upsilon_B$ to underscore the contrast between tracing the RGB and tracing young, main sequence stars at optical wavelengths, although a more symbolic comparison may be with the I-band M/L.  The I-band is preferred to the shorter wavelength B-band for tracing the stellar mass for the same reasons we prefer 3.6 $\mu m$ imaging of old stars \citep{bdJ01} (reduced sensitivity to young stars and dust extinction).  It has even been argued that the I-band is favorable to NIR (and 3.6 $\mu m$) bands for estimating stellar mass, given the uncertain contribution from stars in the late (i.e. AGB) stages of stellar evolution (Rix \& Rieke 1993; Rhoads 1998; Portinari et al. 2004; along with the contribution from non-stellar emission at 3.6 $\mu m$).  Here we emphasize that, for {\it old stellar populations}, we can expect $\Delta\Upsilon_{3.6}$$\sim$0.3$\Delta\Upsilon_{I}$, given their typical colors I-[3.6]=2.2 (I-H=1.8). 
\subsection{Constraining variations with color information}\label{sec:withcolor}
Population synthesis models provide a clean way to explore the dependencies of the observed properties of a stellar population on age/SFH and metallicity, and relate these to the stellar M/L.   
Figure \ref{fig:bothMLgrids} shows the stellar M/L in the optical ($\Upsilon_{B}$) and at 3.6 $\mu m$ ($\Upsilon_{3.6}$) as a function of two different colors for a set of dust-free synthesized stellar populations assuming a Chabrier IMF.  Each population is constructed with an exponentially declining SFH with e-folding time $\tau$ from GALAXEV models (Bruzual \& Charlot 2003; hereafter BC03) and observed after 13 Gyr, or when the population has a mean age A=13 Gyr-$\tau$.

Note first that $\Upsilon_{3.6}$ is characterized by much less variation (a factor of 9 less) than $\Upsilon_{B}$ over this range of ages and metallicities (see also \citealt{dedenus}).  Second, $\Upsilon_{3.6}$ and $\Upsilon_{B}$ show opposite dependencies on metallicity (clearer in the right panel).  The young stars that dominate in the optical are bluer and brighter at lower metallicities (due to a combination of reduced line blanketing and opacity), reducing $\Upsilon_{B}$ compared to higher metallicities.  The metallicity dependence of $\Upsilon_{3.6}$ is reversed, since the RGB traced in the NIR is brighter and redder at higher metallicities.  

These variations in metallicity are well-tracked by the optical-NIR color, which exhibits a weaker dependence on age than the B-R color.  In the left panel, we see that age and $Z$ contribute more equally to the B-R color and introduce comparable variation in $\Upsilon_{3.6}$, while their individual influences on $\Upsilon_{B}$ are degenerate (Bell \& de Jong 2001).  
The R-H color breaks this degeneracy, revealing similar degrees of variation in $\Upsilon_{B}$ due to age and metallicity over the ranges in these parameters considered here.  The nearly orthogonal dependencies of $\Upsilon_{3.6}$ on age and metallicity are likewise responsible for similar degrees of variation in $\Upsilon_{3.6}$.  But, again, these are about one-tenth as large as in the B-band.  

These well-known dependencies of optical and optical/NIR colors on population age and metallicity offer the opportunity to estimate the stellar M/L: the B-R color provides a good constraint on $\Upsilon_{B}$ \citep{bdJ01}, while the combination of R-H and B-R provide the potential to estimate $\Upsilon_{3.6}$ (by constraining the metallicity and the age, respectively; as used by \citealt{z09}).  But such determinations are at risk of the uncertainties introduced by dust (explored below), to which colors tied to optical bands, in particular, are sensitive.  

\subsection{Impact of dust}\label{sec:dust}
The task of assigning M/L requires finding and using a color (or colors) to optimally trace these age and metallicity variations.  Unfortunately, this task is complicated by the presence of dust, which can reduce the diagnostic power of colors tied to optical bands in disk galaxies, in particular (\citealt{bdJ01}; \citealt{macarthur}; Z09).  

The impact of dust at optical wavelengths is demonstrated in Figure \ref{fig:bothMLgrids} by the extinction vector A$_V$=0.3 at the top left, assuming a Galactic extinction law (\citealt{cardelli}; \citealt{indebetouw}).   Dust both reddens and dims (optical) light, increasing B-R and $\Upsilon_{B}$.  But because the dust vector is not exactly parallel to the locus of dust-free stellar population models, the broken age/metallicity-dust degeneracy leads to underestimated masses for young, dust-reddened populations relative to old, dust-free populations.
As a result of dust, typical uncertainties on total masses are 0.1-0.2 dex using global colors to assign the stellar M/L (Bell \& de Jong 2001), although colors are expected to remain reliable in 1D for all but the most massive (and therefore dustiest; i.e. Dalcanton \& Bernstein 2002) disks (MacArthur et al. 2004).  As our main concern is not only the total mass, but the 2D mass distribution within the disk, dust extinction can pose a real problem.  This is especially true since we are interested in mapping the stellar mass in dusty, star-forming disks with a wide range of masses.  
In star-forming disk galaxies $A_V$ can locally be as high as 1-3 mag (e.g. \citealt{prescott}; \citealt{kreckel}; \citealt{ganda}), and even higher in edge-on disks (\citealt{jansen}; \citealt{ganda}).   For a Galactic dust extinction law, even $A_V$=1 mag introduces a color excess of 1.5 mag in B-R color, exceeding by far the total amount of variation exhibited by stellar populations with ages between 1 and 12 Gyr.  

Although we can expect optical colors to retain their usefulness, e.g., in early type E/S0 galaxies characterized by lower dust content, for dusty, star-forming galaxies (the bulk of the $S^4$G sample), the use of optical colors clearly requires incorporating a realistic model of the dust (\citealt{macarthur}; Zibetti et al. 2009).  So far, the effect of dust on UV-to-optical emission from late-type spiral disks has been well modeled assuming a mix of foreground screen and birth-cloud extinction (e.g. \citealt{popescu}; \citealt{dacunha}).  Given the variety of dust geometries, which likely vary from position to position in the disk, models can be difficult to implement realistically in 2D without knowledge of the actual dust content and properties measured from FIR data sampling the dust SED.  The first implementation of a dust model for the purposes of estimating the stellar M/L by Zibetti et al. (2009) may not be optimal, as recent work suggests that stellar reddening is a poor tracer of the overall dust distribution \citep{kreckel}.   

\begin{figure}[t]
\includegraphics[width=.9\linewidth]{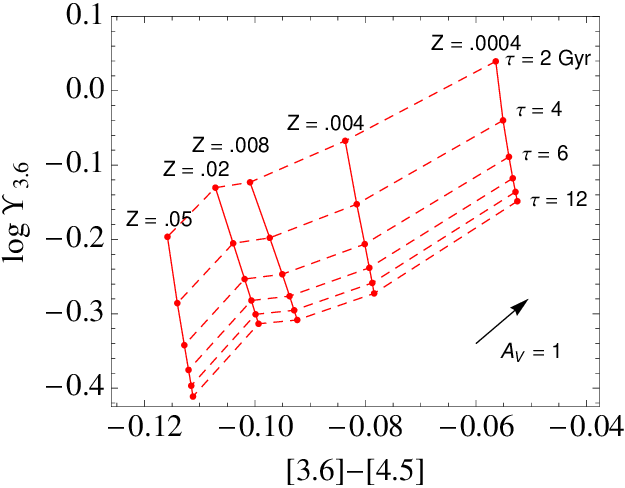}
\caption{$\Upsilon_{3.6}$-([3.6]-[4.5]) relation for a sequence of exponentially declining star formation rate models from Figure 1, using the calibration between J-H and [3.6]-[4.5] in Appendix 1 to link the BC03 models to [3.6]-[4.5] color.  Models of the same e-folding timescale  $\tau$ are connected by dashed lines, while models of the same metallicity Z are connected by soild lines. The black arrow in the bottom right corresponds to V-band extinction A$_V$=1, assuming a Galactic extinction law (\citealt{cardelli}; \citealt{indebetouw}).}
\label{fig:iracMLgrids}
\end{figure}

\begin{figure*}[t]
\begin{centering}
\begin{tabular}{c}
\includegraphics[width=.75\linewidth]{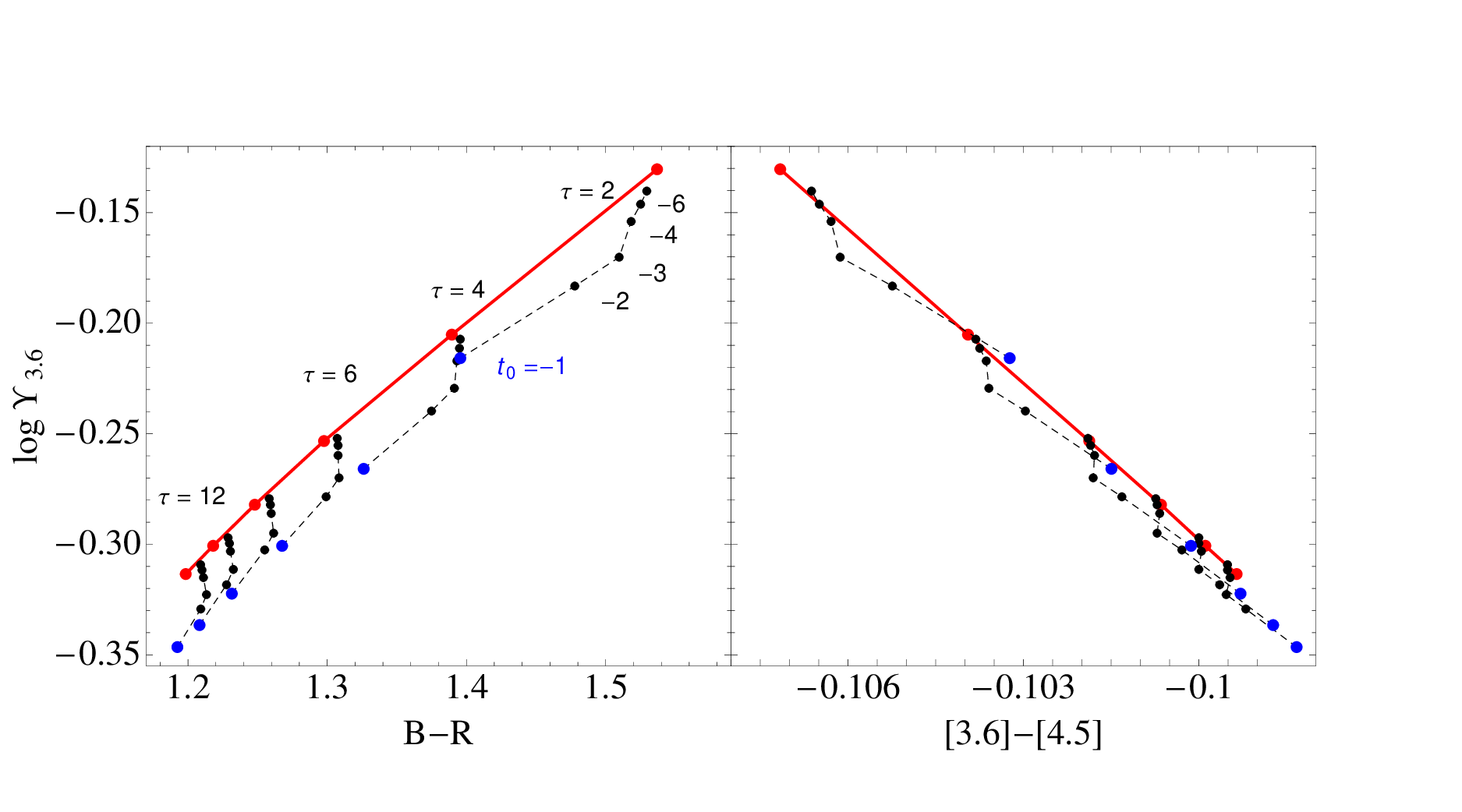}
\end{tabular}
\caption{Mass-to-light ratio $\Upsilon_{3.6}$ (in solar units $M_{\odot}/L_{\odot}$) as a function of B-R and [3.6]-[4.5] color for a sequence of exponentially declining star formation rate models at solar metallicity (Z=0.02) 
with 10\% mass fraction added in 0.5 Gyr-duration starbursts. The solid red line connects the exponential SFH models with different e-folding timescales $\tau$ from Figure \ref{fig:iracMLgrids}. The dashed lines connect models of the same $\tau$ with starbursts occurring 1, 2, 3, 4, or 6 Gyr ago.  The SFH with the youngest mean age (latest burst) is shown in blue for each underlying decaying exponential. }
\label{fig:iracMLbursts}
\end{centering}
\end{figure*}

\section{Constraining $\Upsilon_{3.6}$ with [3.6]-[4.5] colors}\label{sec:iraccolor}
In what follows we describe an approach to estimating $\Upsilon_{3.6}$ that attempts to avoid the effects of dust altogether, taking advantage of the reduced sensitivity of the NIR window to dust extinction.  We explore how well the [3.6]-[4.5] color, in particular, can constrain $\Upsilon_{3.6}$ and how this can be applied to the bulk of S4G galaxies.  

Throughout, we assume that 3.6 $\mu m$ luminosities predicted by the models are accurate, at least over the ages ($>$1  Gyr) under consideration.  Unlike the 4.5 $\mu m$ band that contains a CO absorption feature which is either omitted (BC03) or for which the modeled metallicity dependence is questionable (typically scaled solar), there is no significant photospheric molecular absorption in the 3.6 $\mu m$ band.  So although the modeled 4.5 $\mu m$ luminosities, and thus [3.6]-[4.5] colors, are compromised, the 3.6 $\mu m$ luminosity should remain accurate for the ages under consideration.  Only at the onset of the AGB phase (ages $\lesssim$1Gyr), are modeled 3.6 $\mu m$ luminosities potentially unrealistic.  

To convert from 3.6 $\mu m$ luminosity to magnitude we use the IRAC
channel 1 zero-point calibration provided by Reach et al. (2005) (280.9 Jy; for channel 2 the zero-point is 179.7 Jy).  We adopt the 3.6 $\mu m$ and 4.5 $\mu m$ absolute solar magnitudes calculated by \cite{oh}, $M_{\odot}^{3.6}$=3.24 and $M_{\odot}^{4.5}$=3.27.  All mass-to-light ratios are presented in solar units $M_{\odot}/L_{\odot}$ and all magnitudes and colors are in the Vega magnitude system.

\subsection{Trends with [3.6]-[4.5] color}\label{sec:iraccolor}
\indent In this section, we explore the specific age, SFH and metallicity dependences of the [3.6]-[4.5] colors of old stellar populations. The colors of red giant branch stars may not provide the leverage on age/SFH that the (optical) colors of main sequence stars can.  Even so, between 3.6 and 4.5 $\mu m$ these colors are less susceptible to the systematic uncertainties plaguing optical colors (e.g. dust extinction/reddening) otherwise used to constrain the M/L.  Plus, any variations in M/L due to age will be intrinsically much less than at shorter wavelengths.  In this case, the [3.6]-[4.5] colors of old stars may alone provide suitable constraints on the M/L.  

\indent Unfortunately, SPS models are so far not well-trusted at IRAC wavelengths, where molecular line opacities and the TP-AGB phase of evolution remain weakly constrained in template spectra.  Peletier et al. (2012) show the degree to which predictions vary from model-to-model.  We note that the modeled [3.6]-[4.5] colors do not match the observed global colors of galaxies, either E/S0 (Peletier et al. 2012) or later-type \citep{pahre}, which do not exceed [3.6]-[4.5]=0 and strictly fall in the range -0.15$<$[3.5]-[4.5]$<$-0.02.  Models that do not properly include CO absorption in the 4.5 $\mu m$ wavelength range tend to predict redder [3.6]-[4.5] colors (BC03 and CB07; see \ref{fig:modvbb88}) or a weaker dependence on metallicity (\citealt{marigo}; Maraston 2005) than observed.  \footnote{Dust emission reddens [3.6]-[4.5] and, as shown in Paper I, this is responsible for the distribution of observed [3.6]-[4.5] colors in nearby galaxies varying between -0.01 and $\sim$ 0.3.}  Even at ages long after the AGB phase, the models have very little predictive power at mid-IR wavelengths.  To avoid this issue, in Appendix 1 we calibrate a relation between the NIR and [3.6]-[4.5] colors of GLIMPSE giants in order to relate the NIR colors of SPS models -- which successfully reproduce the colors of real galaxies -- to realistic [3.6]-[4.5] colors.  These better match expectations (based on observation) for the mid-IR colors of old stellar populations.  

\indent Figure \ref{fig:iracMLgrids} shows the same BC03 tracks of metallicity and age as shown in the previous section, but now using the relation defined in Appendix A.  Notably, the trend in $\Upsilon_{3.6}$ with [3.6]-[4.5] is the reverse of the trend with R-H color in Figure \ref{fig:bothMLgrids};   
due to CO absorption in the stellar photospheres of giant stars in the 4.5 $\mu m$ band, the [3.6]-[4.5] color is bluer at higher metallicity.  
In addition, 
only the influence of metallicity can account for the full range in [3.6]-[4.5] colors observed in real galaxies (as measured, e.g., by \citealt{pahre}).  
Peletier et al. (2012) find this same dependence of [3.6]-[4.5] on metallicity in early-type galaxies, linking the relation between [3.6]-[4.5] and galaxy velocity dispersions in the SAURON sample to a mass-metallicity relation, and also linking [3.6]-[4.5] directly to absorption lines like Mgb and Fe.     
With our GLIMPSE-based calibration to relate [3.6]-[4.5] color to M/L it is now clear that metallicity is in fact the primary driver of [3.6]-[4.5], the age dependence of [3.6]-[4.5] being weak.  Indeed, age variation introduces a limited span of only $\sim$0.002 mag in [3.6]-[4.5] color from $\tau$=2-12 Gyr at fixed metallicity (see Figure \ref{fig:iracMLgrids}).   
Variation in metallicity across the range 0.0004$<$Z/Z$_{\odot}$$<$0.05, on the other hand, can account for colors spanning -0.08$\lesssim$[3.6]-[4.5]$<$-0.02.  

Over the full range in [3.6]-[4.5] probed here, metallicity introduces a change in M/L by 0.2 dex.  At fixed metallicity, age variations are responsible for a similar degree of variation, but the corresponding change in [3.6]-[4.5] color of only 0.002 mag is below the typical photometric accuracy of these bands (Reach et al. 2005).  As is the case for NIR colors to which our IRAC color is tied, age variations are not well probed.  Still, as discussed in $\S$ \ref{sec:MLdescription}, age variations themselves do not lead to greatly varying $\Upsilon_{3.6}$.  In fact, Figure \ref{fig:iracMLgrids} suggests that these are at most $\pm$ 0.1 dex at fixed metallicity.  Considering the range of ages and metallicities exhibited by the galaxies studied by \citet{macarthur}, the oldest age (lowest $\tau$) considered here may even be unrealistic.  More realistic SFHs,  characterized by at least one additional burst of star formation during the decline of initial star formation, correspond to younger mean ages and, as demonstrated in the next section, tend to tighten the spread in $\Upsilon_{3.6}$ at fixed metallicity.  

\subsection{The influence of (bursty) SFHs}\label{sec:bursts}
Here we explore the impact of bursts on $\Upsilon_{3.6}$.  For each of the solar metallicity (Z=0.02) exponentially declining SFHs we add one burst of star formation with 0.5 Gyr duration.  Each burst contributes 10\% to the total stellar mass formed over the lifetime of the population and is chosen to occur between 1 and 6 Gyr from the present `observation' epoch.  Following MacArthur et al. (2004), we assume that more recent and more massive bursts are not likely in our sample, according to the Kauffman et al. (2003) study of burst fraction in the Sloan Digital Sky Survey.  Note that this simple model does not account for bursts that produce stars with metallicities other than the metallicity characteristic at earlier times.  

As can be seen in Figure \ref{fig:iracMLbursts}, the [3.6]-[4.5] color is less sensitive to bursty SFHs than the B-R color, as expected (i.e. the [3.6]-[4.5] color itself is not very sensitive to age; see the previous section).  However, older bursts introduce a slightly larger fractional change to the [3.6]-[4.5] color than to the B-R color.  This change in [3.6]-[4.5] stays in roughly fixed proportion to the change in $\Upsilon_{3.6}$ for bursts of all ages.  As a result, the fixed metallicity age track in the right panel remains robust; with the addition of bursts the population slides down the righthand side age track, while on the left the age track itself is changed.  

In either the left or right panels of Figure \ref{fig:iracMLbursts}, it is clear that later bursts have the largest impact on color and $\Upsilon_{3.6}$, and this increases for shorter $\tau$.  Imposing a burst  of star formation on an exponentially declining SFH primarily reduces the mean age of the stellar population (i.e. MacArthur et al. 2004) and thereby also effectively reduces the spread in $\Upsilon_{3.6}$ compared to the no-burst SFHs.  
Note that 
the [3.6]-[4.5] color for populations with the youngest age (corresponding to a burst 1 Gyr ago) is the most likely to be inaccurate as it does not account for AGB stars (only giant stars).  At $\sim$1 Gyr, the contribution from AGB stars in the NIR is predicted to reach its peak of $\sim$30\% or as high as $\sim$70\%, depending on the model (cf. BC03 and Maraston 2005).  
The trends in Figure \ref{fig:iracMLbursts} described above remain secure for all other SFHs.  

\subsection{The relation between $\Upsilon_{3.6}$ and [3.6]-[4.5] color}\label{sec:therelation}
Given that bursty SFHs reduce the mean population age and hence tighten the spread in $\Upsilon_{3.6}$, metallicity is arguably the source of largest variation in $\Upsilon_{3.6}$.  This is especially clear if we focus only on populations with ages younger than $\sim$10Gyr ($\tau$=4-12 Gyr) in Figure \ref{fig:iracMLgrids}, even discounting the lowest metallicity models.  Such a reduced range of ages and metallicities would be representative of the large majority of nearby S0-Irr galaxies studied by \citet{macarthur}. 

\begin{figure}[t]
\includegraphics[width=.9\linewidth]{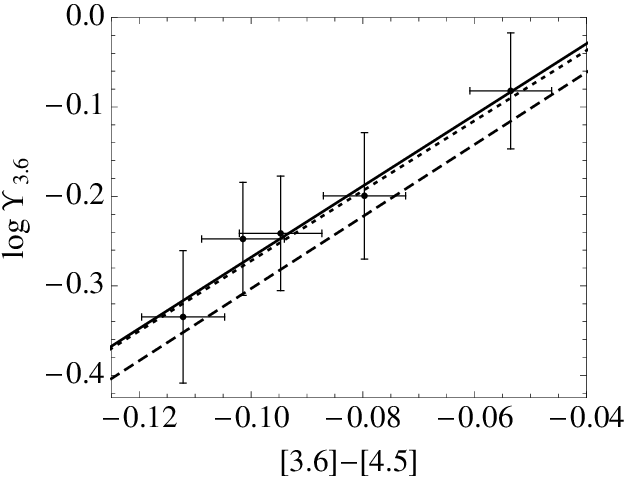}
\caption{Relation between fixed-metallicity bins of $\Upsilon_{3.6}$ and [3.6]-[4.5] measured from the sequence of exponentially declining star formation rate models shown in Figure \ref{fig:iracMLgrids}, with errors as described in the text.  The solid black line shows our best-fit relation.  The dashed and dotted lines show the results of fits over a restricted range of metallicity and age (see Table \ref{tab:fits}).  The bluest point ($Z$=0.0004) has been omitted for the fit traced by the dotted line, while the dashed line shows the fit to slightly different bins (not shown) from which the model with the oldest age at each metallicity is omitted.  }
\label{fig:binning}
\end{figure}

As the [3.6]-[4.5] color is itself primarily driven by metallicity, it serves as a natural constraint on $\Upsilon_{3.6}$.  In Figure \ref{fig:binning} we show our parameterization of the relation between $\Upsilon_{3.6}$ and [3.6]-[4.5].  Assuming that all regions of the parameter space in Figure \ref{fig:iracMLgrids} are equally likely, we bin our set of models by metallicity, letting the spread in $\Upsilon_{3.6}$ and [3.6]-[4.5] due to age set the dispersion around the mean.  On average, unconstrained age variations introduce an rms uncertainty of 0.06 dex on $\Upsilon_{3.6}$, shown as the vertical error bar at fixed metallicity in Figure \ref{fig:binning}.  The horizontal error bar reflects age-induced variation in [3.6]-[4.5] color at fixed metallicity ($\sim$0.002mag) together with the uncertainty in our GLIMPSE-based calibration between J-K and [3.6]-[4.5], with which we can estimate the [3.6]-[4.5] color to within $\sim$0.009 mag (Appendix 1).  

Taking a linear least-squares fit to the points in Figure \ref{fig:binning}, we find 
\begin{equation}
\log{\Upsilon_{3.6}}=3.98(\pm0.98) ([3.6]-[4.5]) + 0.13(\pm0.08). 
\end{equation}
We adopt this fit as our optimal color-M/L relation.  As demonstrated by Table 1, this relation is not especially sensitive to the particular range of age and metallicity probed by our models.  Even with restricted metallicity and age ranges, errors in the slope and intercept primarily quantify the uncertainty introduced by age/SFH.  

Considering that the ages of either early- or late-type galaxies (ETGs and LTGs) are expected to each fall within narrower ranges than the full range considered at each $Z$, the errors on the parameters of our optimal relation between $\Upsilon_{3.6}$ and [3.6]-[4.5] color are even conservative.  Fits only to models in Figure \ref{fig:iracMLgrids} with ages $>$ 7 Gyr ($<$ 7 Gyr), as would be representative of ETGs (LTGs), yield best-fit parameters $a$=3.86$\pm$0.82 and $b$=0.18$\pm$0.07 ($a$=4.07$\pm$0.67 and $b$=0.09$\pm$0.06). 

For completeness we note that the spread in $\Upsilon_{3.6}$ with age at fixed metallicity is larger the nearer the population is observed to the start of star formation.  Also, this relation between $\Upsilon_{3.6}$ and [3.6]-[4.5] does not account for dust extinction.  

In addition, our best-fit relation predicts a change in $\Upsilon_{3.6}$ by 0.2 dex for every 0.05 mag difference in color, slightly larger than typically emerges from standard single (optical)-color relations (see Z09; \citealt{bdJ01}).   We emphasize, though, that by passing from optical colors to NIR-only colors, the stellar tracer changes from MS stars to red giants, and we can expect the latter to show much less variation than the former for the same evolution in a given population.  

\begin{table}[t]
\begin{center}
\caption{Best-fitting relations between $\Upsilon_{3.6}$ and [3.6]-[4.5] color\label{tab:fits}}
\begin{tabular}{rcccc}
\tableline\tableline
 fit type&$a$&$b$&$\delta$$a$&$\delta$$b$\\
\tableline
all&3.98&0.13&0.99&0.08\\
$\tau$&4.03&0.11&0.77&0.06\\
Z&3.92&0.12&1.30&0.12\\
\tableline
\end{tabular}
\tablecomments{\small {Parameters of the best-fit relation $\log{\Upsilon_{3.6}}$=$a$ ([3.6]-[4.5])+ $b$ to the models in Figure \ref{fig:iracMLgrids}, for three different sets of metallicity and age ranges.  Prior to the fit the models are binned at fixed metallicity (e.g. Figure \ref{fig:binning}). Fit-type `all' in the first row corresponds to a fit to all points 0.0004$<$$Z$$<$0.05 and 2$<$$\tau$$<$12 Gyr, with binning as shown in Figure \ref{fig:binning}.  The fit in the second row discards models with the smallest $\tau$ (oldest age). The fit in the third row discards the lowest metallicity $Z$=0.0004, but keeps all ages.  The fit over the restricted range in e-folding $\tau$ (second row) should more realistically depict the trend for galaxies with bursty SFHs (see Figure \ref{fig:iracMLbursts}). }}
\end{center}
\end{table}
\subsection{Color-independent $\Upsilon_{3.6}$}\label{sec:ageZ}
Over the color range -0.14$<$[3.6]-[4.5]$<$-0.04, the 3.6$\mu m$ mass-to-light ratio for the oldest stars spans $\sim$0.3 dex, driven primarily by metallicity.  Using the [3.6]-[4.5] color to constrain the metallicity, the only lingering uncertainty in $\Upsilon_{3.6}$ is due to variation in the population age, which contributes an rms of 0.06-0.08 dex uncertainty at a given metallicity 
(see Figure \ref{fig:binning}).  To reduce the uncertainty in $\Upsilon_{3.6}$ to any greater degree would require better constraints on the population age.  According to the trends described in $\S$ \ref{sec:iraccolor}, this would be possible using an optical color, but only by also incorporating a realistic model of the dust to account for extinction and reddening (as well as the scattering of light by dust).  

On the other hand, we can leverage the age-metallicity relation (AMR) characteristic of old stellar populations to place a constraint on the age at a given metallicity.  With the [3.6]-[4.5] color tracing both the metallicity and the age, the uncertainty in $\Upsilon_{3.6}$ can be reduced.  The age-metallicity relation even suggests that a constant $\Upsilon_{3.6}$ may fairly reasonable, given that the relation between $\Upsilon_{3.6}$ and $Z$ is almost orthogonal to the relation between $\Upsilon_{3.6}$ and $\tau$.  As such, old stellar populations across the Hubble sequence seem to be well-described with a constant $\Upsilon_{3.6}$.  Indeed, according to Figure \ref{fig:iracMLgrids}, old, metal-rich populations have comparable $\Upsilon_{3.6}$ to younger, more metal-poor populations (compare $\Upsilon_{3.6}$ at $Z$=0.05, $\tau$=4Gyr with $Z$=0.004, $\tau$=12Gyr). 

We confirm the implied independence of $\Upsilon_{3.6}$ on color 
by adopting the relation between the ages and metallicities of elliptical galaxies measured by \cite{gallazi1} across the stellar mass range 8.91$<$$\log{M_*/M_\odot}$$<$11.91.  This supplies an estimate of the age at a given assumed metallicity, which we then translate back to the quantities $\Upsilon_{3.6}$ and [3.6]-[4.5] color using metallicity-dependent linear fits of the relation between each quantity and population age from the set of models in Figure \ref{fig:iracMLgrids}.  
Over the range of \cite{gallazi1} measured metallicities (from 0.002$\lesssim$$Z$$\lesssim$0.02), $\Upsilon_{3.6}$ exhibits much less variation with color than implied in Figure \ref{fig:iracMLgrids}.  The age-metallicity relation is consistent with a constant $\Upsilon_{3.6}^{AMR}$=0.6 for a Chabrier IMF\footnote{With a Salpeter IMF this corresponds to $\Upsilon_{3.6}$=1.07.}, good to within 0.06 dex.  The uncertainty stems from the dispersion around the mean value over the adopted metallicity range, together with propagated errors due to the uncertainty in the fitted AMR represented by the intrinsic 1-sigma spread in the \cite{gallazi1} measurements.   

The relation between old stellar age and metallicity allows us to know $\Upsilon_{3.6}$ even better than suggested by the full grid of models in Figure 2.  Uniformly weighted, those models are consistent with an average constant (color-independent) $\Upsilon_{3.6}$=0.6 (Chabrier) and an estimated uncertainty 0.11 dex.  But while our $\Upsilon_{3.6}^{AMR}$ incorporates a realistic intrinsic spread in the relation between age and metallicity as measured by \cite{gallazi1}, it should be used with caution.    

The existence of an age-metallicity relation in stellar populations depends on the history of star formation in these systems.  We have assumed that the age-metallicity relation measured for the old stellar populations in elliptical galaxies can be applied to the old stellar populations in late-type galaxies (isolated from younger populations at 3.6 $\mu m$).  This  implicitly assumes that late-type disks had similar formation and evolutionary histories as present day ellipticals, up to the point when the latter underwent quenching (and an associated morphological change) whereas the former continued to form stars.  

For systems in which an AMR in the stellar populations is not likely (e.g. due to an atypical evolutionary history), we therefore recommend either adopting the more conservative uncertainty of 0.11 dex on the constant $\Upsilon_{3.6}$=0.6, or using the [3.6]-[4.5] color-dependent relation in $\S$ \ref{sec:therelation}, which is preferred when color information is available.  

The same applies locally, within stellar systems, given indications that a link between age and metallicity persists within disks (e.g. \citealt{macarthur}).  This implies that $\Upsilon_{3.6}$ will be relatively flat within a galaxy with population gradients (as well as that a single $\Upsilon_{3.6}$ can be applied to galaxies of all types).  The observed 2D [3.6]-[4.5] color distribution can be used to assess which $\Upsilon_{3.6}$ and uncertainty may be most appropriate.  An implied smooth, monotonic variation in metallicity ([3.6]-[4.5] color) throughout the system may indicate that $\Upsilon_{3.6}^{AMR}$ can be reliably applied; otherwise, the color-dependent relation should be adopted. In cases where color information is compromised (either because of contamination from non-stellar emission or data quality), we advocate the use of the most conservative estimate, $\log{\Upsilon_{3.6}}$=-0.22$\pm$0.11 dex 

The remarkable single, color-independent $\Upsilon_{3.6}$ implied by the existence of an AMR within (and among) galaxies turns out to be a particular advantage for generating mass maps from our ICA maps of the old stellar light.  As noted in $\S$ \ref{sec:sources}, it may not be possible to retain full 2D color information tracking age and/or metallicity gradients in galaxies with pervasive non-stellar emission \citep{meidtpaperI}.  The old stellar light map can nevertheless be converted to stellar mass with $\sim$15 \% accuracy (adopting the single, global [3.6]-[4.5] color that is retained; Querejeta et al. 2014).

\section{Comparison with alternative approaches}\label{sec:altApproaches}
In the previous section we defined a relation between $\Upsilon_{3.6}$ and [3.6]-[4.5] for the oldest stars, the scatter about which is driven by age variation.  At the same time, the age-metallicity relation may favor a color-independent $\Upsilon_{3.6}$.  Together, these suggest that $\Upsilon_{3.6}$ can be estimated to within $\sim$0.06-0.1 dex ($\sim$15-30\%; depending on whether $\Upsilon_{3.6}$ is taken to depend on [3.6]-[4.5] color).  Remarkably, the uncertainty introduced by unconstrained age variations greatly improves upon the typical uncertainties associated with stellar M/Ls at shorter wavelengths using a single (optical-NIR) color ($\sim$0.2 dex; \citealt{bdJ01}).  Our uncertainty is even comparable to that achieved by using two colors to assign the M/L ($\sim$0.05-0.1 dex; Zibetti et al. 2009).  Moving to longer wavelengths and using only one color defined between 3.6 and 4.5 $\mu m$ already achieves an accuracy similar to that which can be obtained with state-of-the-art techniques that use double the information (and which rely on dust models that may themselves not be entirely realistic).  Below we compare our own $\Upsilon_{3.6}$ to several other estimates specific to the 3.6 $\mu m$ band. 
\subsection{LMC-based calibration of $\Upsilon_{3.6}$ }
Recently, \citet{eskew} assembled an estimate of $\Upsilon_{3.6}$ by using resolved measurements of the  SFH of the LMC \citep{HZ} to measure the mass in regions sampling throughout the map of 3.6 $\mu m$ brightness.  They find that mass is well traced by 3.6 $\mu m$ flux and, based on the relation between the two, estimate $\Upsilon_{3.6}$=0.53, with an uncertainty of $\sim$30\%.  Then, finding that the scatter about the relation tends to correlate with the [3.6]-[4.5] color, they fit the  additional relation
\begin{equation}
\log{\Upsilon_{3.6}}=-0.74 ( [3.6]-[4.5] ) -0.23
\end{equation}
attributing some part of the trend with [3.6]-[4.5] to the presence of either younger stellar populations ($<$1Gyr) or dust emission (traced at 8$\mu m$).  

These estimates are reassuringly consistent with our own, although the second, in terms of [3.6]-[4.5] color, suggests a different (much weaker) color dependence than we found in $\S$ \ref{sec:iraccolor}.    
This may be expected, given that the fairly uniform metallicity of the LMC would restrict the range in [3.6]-[4.5] color. 
But if all of the color variation stems from age variation in this case (ages range from 12 Gyr to $<$1 Gyr in the LMC; \citealt{HZ}), then the shallower color dependence may suggest either that $\Upsilon_{3.6}$ is weakly dependent on young populations, in particular, or that dust emission acts to remove the steeper trend we find.  
\subsection{$\Upsilon_{3.6}$ in the presence of non-stellar emission}

There are several techniques to estimate the 3.6 $\mu m$ mass-to-light ratio in the presence of non-stellar emission, $\Upsilon_{3.6}^d$, based on global colors and total masses.  \citet{zhu} link SDSS masses with Spitzer luminosities and colors in SWIRE.  \citet{jarrett2013} use the \citet{zhu} 2MASS relations to derive K-band masses and link these to WISE luminosities and colors for a small sample.  Both relations extend redder than [3.6]-[4.5]=0 and parameterize a dependence on [3.6]-[4.5] that is nearly the reverse of what we find.  \citet{zhu} also suggest that $\Upsilon_{3.6}^d$ is much more sensitive to SFH than metallicity, contrary to our findings.  

These calibrations primarily reflect the fact that, as the non-stellar emission tracing young stellar populations makes an increased fractional contribution to the integrated 3.6 and 4.5 $\mu m$, the global [3.6]-[4.5] color reddens (even independent of the dust properties) and the total luminosity increases, thus reducing the mass-to-light ratio.  

The fractional contribution of non-stellar emission to the integrated light of a galaxy can be increased due to, e.g., an enhanced star formation rate (at fixed mass) or an intrinsically less massive stellar component (at fixed SFR).  Both of these cases would be associated with redder [3.6]-[4.5] colors, and thus a lower M/L.  But at the local level, the [3.6]-[4.5] color in galaxies also reflects real changes in the dust properties (PAH vs. larger grains and the physical state of the dust; see Paper I), rendering the above estimates ineffective.   For mapping the 2D stellar mass distribution, we advocate first removing the non-stellar emission and then applying the $\Upsilon_{3.6}$ derived in this work.  

\section{Discussion}   
\subsection{Stellar population models \& IMF}
\indent The relation between $\Upsilon_{3.6}$ and [3.6]-[4.5] color presented here, while robust, still hinges on the assumed stellar population model and IMF.   
The J-H colors of old stellar populations, to which our [3.6]-[4.5] colors are tied, are fairly consistent from model to model (Charlot \& Bruzual 2007), even as predictions for optical-NIR colors like V-K vary by  $\sim$0.3 mag at fixed age.  On the other hand, discrepancies in the NIR M/L of populations with (exponentially) declining SFHs emerge at all ages, as a result of differences in the treatment of stars in the AGB phase that occurs at SSP ages $\sim$ 1 Gyr. Depending on the model, predictions for the NIR contribution of AGB stars at this intermediate age are as high as 70\% (Maraston 2005; Charlot \& Bruzual 2007, hereafter CB07) or as low as 30\% (BC03).  As a result, $\Upsilon_H$ predicted by CB07 and BC03 differ by $\sim$0.1 dex at all ages \citep{z09}, with CB07 predicting a lower M/L.\footnote{In synthesized populations with continual star formation the asymptotic giant branch is continually populated, but the stars there are dominated by main sequence stars at all but the youngest ages.  At old ages, the variation from BC03 to CB07 remains minor (and nearly uniform), given the predominance of main sequence stars over AGB stars in the population (i.e. Zibetti et al. 2009).} 
The larger NIR contribution of AGB stars at intermediate ages predicted by the \citet{maraston05} models would imply a similarly lowered M/L.  In these cases, $\Upsilon_{3.6}$ would be reduced by $\sim$ 0.1 dex across all [3.6]-[4.5] colors.  

We have opted to use the BC03 models for this work, as recent evidence suggest that the assumed AGB prescription may be the most realistic.  The NIR spectra of post-starburst galaxies studied by Zibetti et al. (2012) do not exhibit the strong molecular absorption predicted by \citet{maraston05} models.   A lower NIR contribution from AGB stars, as in the BC03 models, seems to provide the best fit to the SEDs of post-starburst galaxies analyzed by Kriek et al. (2010).  A similarly low contribution is inferred for AGB stars in $\sim$1Gyr old clusters in M100, arguably due to the extinction of the AGB by the circumstellar dust shell produced during this phase (\citealt{meidt2012b}; which leads to reprocessed emission detected at mid-IR wavelengths).  We therefore argue that the relation we derive based on BC03 models is optimal, and that the uncertainties stemming from the properties of stellar populations are representative.  

\indent 
Only the uncertainty arising with the assumed IMF will dominate the lingering uncertainty in $\Upsilon_{3.6}$ due to age/SFH.  The Chabrier IMF, for one, entails an increased fraction of giants to dwarfs at low turn-off mass $M_{TO}$, effectively decreasing the total mass-to-light ratio below the traditional Salpeter-based value (although the heavier main sequence will partially compensate in this case).  Generally, the choice of the IMF will introduce systematic variation in $\log{\Upsilon_{3.6}}$ by 0.05 dex to 0.3 dex (see \citealt{bdJ01}; Z09; \citealt{bernardi}).  Given that the uncertainties on $\Upsilon_{3.6}$ associated with the IMF exceed those arising with the stellar population, it should be possible to use $\Upsilon_{3.6}$ to explore variation in the IMF, e.g., with global galaxy parameters.  

\subsection{Dust emission}
The presence of non-stellar emission is another potential source of discrepancy.  We can expect the (positive) trend between $\Upsilon_{3.6}$ and [3.6]-[4.5] color to weaken or even reverse \citep{jarrett2013} when dust contributes to the [3.6]-[4.5] color and the mass-to-light ratio is therefore $\Upsilon_{3.6}^{d}$=$M/(L_{3.6}^{dust}+L_{3.6}^{stars})$.  In such cases, our strictly stellar $\log{\Upsilon_{3.6}}$=-0.23$\pm$0.1 dex can still be useful and can even provide an estimate for the fractional contamination by dust emission, since 
\begin{equation}
\frac{L_{3.6}^{dust}}{L_{3.6}^{dust}+L_{3.6}^{stars}}=1-\frac{\Upsilon_{3.6}^{d}}{\Upsilon_{3.6}}\\
\end{equation}
\newline
where, i.e., $\Upsilon_{3.6}^{d}$ is estimated with the observed dust reddened [3.6]-[4.5] following \citet{jarrett2013}.\\
\indent We emphasize, again, that our estimate for $\Upsilon_{3.6}$ here is most appropriate for 2D maps where the non-stellar emission has been removed, and which thus arguably trace a uniformly old stellar population. 
Uncorrected 3.6 $\mu m$ maps would most certainly require a more highly color-dependent M/L (at least locally) given the contribution from non-stellar emission.  Without removing such emission, which can locally contribute as much as 30\% (Paper I), then applying our color-independent stellar $\Upsilon_{3.6}$=0.6 will result in significant over-densities, e.g. at the locations of dusty spiral arms.  
The overestimation in 2D stellar mass surface density exceeds the uncertainty associated with our optimal estimate of $\Upsilon_{3.6}$.\footnote{The errors associated with the presence of non-stellar contaminant emission are comparable to the 0.1-0.25 dex uncertainty in stellar M/L estimates based on single-color M/L relations or the 0.1-0.4 dex uncertainty owing to the choice of the AGB prescription or IMF adopted in the models (Z09).}  
The local effect of the contaminants can be quite large and should be kept in mind, e.g., when estimating the stellar mass in individual structural components mapped at 3.6 $\mu m$.  The global contribution from dust will be smaller, introducing $\sim$10\% overestimation in the total mass (Paper I).  

\subsection{Future advances}\label{sec:future}
In $\S$ \ref{sec:ageZ} we discussed the use of the age-metallicity relation in old stellar populations to reduce the uncertainty in $\Upsilon_{3.6}$ due to age variations left unconstrained by the [3.6]-[4.5] color, as in Figure \ref{fig:binning}.  In the alternative, age constraints are possible using optical colors dominated by younger stars, with the incorporation of a realistic dust model to account for extinction and reddening (as well as the scattering of light by dust; e.g. \citealt{z09}).  
Ideally, some part of the constraint on the dust model would come from the FIR dust emission itself.  SED fitting with MAGPHYS (da Cunha et al. 2011), which samples throughout the combined dust and stellar SED, can provide a dust-free view of the properties of the stellar population, and is potentially very powerful when applied pixel-by-pixel.  

On the other hand, it might be possible to use the information on the dust emission contained in the IRAC bands (which can be isolated with ICA).  As this is primarily emission from the smallest grains (PAH and the hottest dust), it would be necessary to first define a relation between these grains and the bulk of the dust.  
\section{Summary and Conclusions}
We have shown that the information contained in IRAC images at 3.6 and 4.5 $\mu m$ is optimal for estimating stellar masses of nearby galaxies.  At 3.6 $\mu m$ we combine an intrinsically modestly varying stellar M/L with the ability to constrain variations in the properties of the stellar population via the [3.6]-[4.5] colors of old stars.  We achieve an uncertainty as low as 0.06 dex by using the [3.6]-[4.5] color to constrain the metallicity-dependence of the stellar M/L; more modest unconstrained age variations at these long wavelengths contribute an uncertainty that is comparable to, if not below, that which can be achieved at shorter wavelengths, where the effects of dust extinction/reddening and young stars are present.  Much of the uncertainty normally introduced with the use of SPS models at long wavelengths ($\lambda\gtrsim$ 2.2 $\mu m$) is reduced by using the observed colors of GLIMPSE giants to relate the NIR colors of SPS models to realistic [3.6]-[4.5] colors.   

Our preferred method of assigning the stellar M/L uses the [3.6]-[4.5] color to account for the increase in the brightness of giant stars with increasing metallicity (traced by [3.6]-[4.5] color).  We define a relation between [3.6]-[4.5] color and $\Upsilon_{3.6}$ that extends across the full range in [3.6]-[4.5] colors exhibited by galaxies across the Hubble sequence.  

This relation is designed to be applied to the light from old stars, alone, i.e. when [3.6]-[4.5] colors reflect genuine variation in the properties of the old stellar population.  This makes it applicable to early-type galaxies, with little on-going star formation (and minimal dust extinction/reddening), or in late-type galaxies, where non-stellar contaminating emission is isolated from the old stellar light with ICA (Paper I).  

Remarkably, the existence of an age-metallicity relation in old stellar populations may even lead to a preference for a single, color-independent $\Upsilon_{3.6}$=0.6, good to an accuracy of 0.06 dex. We recommend adopting a more conservative 0.1 dex uncertainty on $\Upsilon_{3.6}$=0.6 given a suspected atypical evolutionary history, or when it is not possible to correct for the presence of non-stellar emission (which leads to characteristically red [3.6]-[4.5] colors).  
In this case, additional uncertainty in the stellar mass estimate can arise as a result of the contribution from the non-stellar emission, which can be as large as 30\% locally (i.e. an uncertainty comparable to that associated with the stellar M/L).   

Either of the two estimates for $\Upsilon_{3.6}$ presented here (with or without [3.6]-[4.5] color information) mark a significant improvement in our view of how the stellar mass is distributed in and among galaxies.  With greater confidence in the stellar M/L we also gain much-needed leverage on the dark matter content of galaxies, as well as the opportunity to explore variations in the IMF with global galaxy properties.  
We anticipate successfully implementing our preferred approach, one that yields the smallest uncertainty in stellar mass, to galaxies in S$^4$G; future work will first remove the contribution from non-stellar emission in S$^4$G 3.6 $\mu m$ images with ICA and then use the [3.6]-[4.5] color of the old stellar light to assign $\Upsilon_{3.6}$ (Querejeta et al. 2014).    \\
\newline
\newline
Thanks to Mariya Lyubenova for fruitful discussion and the entire S$^4$G team.  E.A., A.B., J.K., G.vdV., M.Q., S.M. and E.S. acknowledge
financial support of the DAGAL network from the People
Programme  (Marie Curie Actions) of the European Union's Seventh
Framework Programme FP7/2007-2013/ under REA grant agreement number
PITN-GA-2011-289313.  K.S., J.-C.M.-M., and T.K. acknowledge support from the National Radio Astronomy Observatory, which is a facility of the National Science Foundation operated under cooperative agreement by Associated Universities, Inc.
 \appendix\section{Appendix A: IRAC colors of giant stars}
The GLIMPSE survey of the galactic mid-plane (\citealt{benjamin}; \citealt{churchwell}), which combines 2MASS JHK with Spitzer/IRAC imaging observations, furnishes the opportunity to calibrate an empirical relation between the NIR colors of giant stars and their [3.6]-[4.5] colors (specific to the IRAC bands).   
From all catalogued GLIMPSE objects detected simultaneously in the J, H, K, 3.6 and 4.5 micron bands brighter than apparent $m_{3.6}$=11 mag, we select candidate giant stars by initially imposing the following criteria:
\begin{eqnarray}
0.7 < \mbox{J-K$_s$} &<&  1.3 \\
0.1 <  \mbox{H-K$_s$} &< & 0.4\\
\mbox{[3.6]-[4.5]} &< & 0.1 \\
\mbox{[5.8]-[8]} &< & 0.1
\end{eqnarray}
The nearest dwarf stars that match these criteria have sufficiently different spectral properties in the NIR at fixed V-K color (bluer J-H colors; Bessell \& Brett 1988, hereafter BB88) that our classification criteria (described below) should naturally select against them.  But to prevent contamination from stars reddened by foreground dust, as well as `infrared excess' sources with non-photospheric dust emission, we impose an additional requirement K-[8]$<$0.1 (Origlia 2010; \citealt{indebetouw}).  
This limit to reprocessed dust emission (re-radiated at longer wavelengths) also helps avoid mis-classifications of dust-reddened giants (Origlia 2010).  

To classify the giants we impose three constraints per spectral type that must be simultaneously satisfied.  The giant spectral type is defined over a narrow bin $\pm$0.05 mag in width around each of three catalogued NIR colors (J-H, H-K and J-K) in the classification by BB88, adopting the transformation to 2MASS colors defined by \cite{carpenter}.  
The bin width is set to the average photometric color error reported in the GLIMPSE catalog for all 3 colors, which is also the typical color difference between BB88 spectral types.  
By imposing these strict constraints in combination with our limit on reprocessed dust emission, we aim to select against dust-reddened stars that appear to match the typical NIR colors of giants.  

\begin{figure}[t]
\includegraphics[width=.5\linewidth]{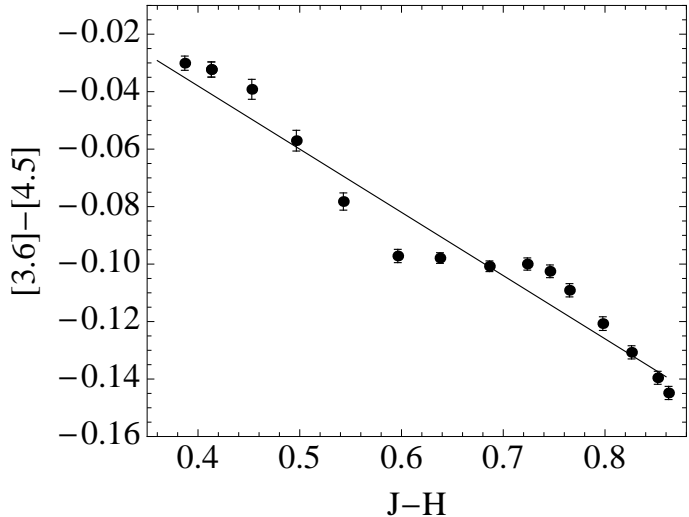}
\caption{The mean [3.6]-[4.5] color and its error for each of the 15 spectral classes against the J-H color of each class of giants. Each bin contains hundreds of stars.  
The solid black line shows our best-fit linear relation between the [3.6]-[4.5] and J-H colors of GLIMPSE giants sorted by spectral type.   (Table \ref{tab:glimpsefits}). }
\label{fig:glimpseCal}
\end{figure}

Each spectral type (defined according to temperature) contains a mix of ages and metallicities.  
Because we are sampling the Rayleigh-Jeans tail, the relations between the NIR colors of giants remain robust to changes in metallicity, keeping the ratio of colors nearly independent of color.  As demonstrated in Figure \ref{fig:glimpseCal}, this also appears to remain true for the mid-IR [3.6]-[4.5] color in relation to any given NIR color, given the fact that it tracks CO absorption in the 4.5 $\mu m$ band (and therefore reflects more than just a temperature dependence on metallicity).  
Although there is some non-linearity,\footnote{The plateau in [3.6]-[4.5] color in the range 0.55$<$J-H$<$0.75 (spectral class K0-K5; Figure \ref{fig:glimpseCal}) may reflect the genuine behavior of CO absorption, as it echos the trend in the equivalent width of the $^{12}$CO absorption band at $\lambda$=2.29 $\mu m$ with giant spectral class demonstrated by \citet{cesetti}. But since the (photometric) [3.6]-[4.5] color uncertainty is on the order of 0.02 mag for these GLIMPSE objects, this behavior corresponds to only modestly significant departures in [3.6]-[4.5] color away from our adopted linear relation.   } 
we can fit the 15 total spectral classes with the following linear relation:
\begin{equation}
[3.6]-[4.5]= -0.22 (J-H) + 0.05
\end{equation}
Table \ref{tab:glimpsefits} lists the best-fit linear relations between other NIR colors and [3.6]-[4.5] color.  

\begin{figure}[t]
\includegraphics[width=.55\linewidth]{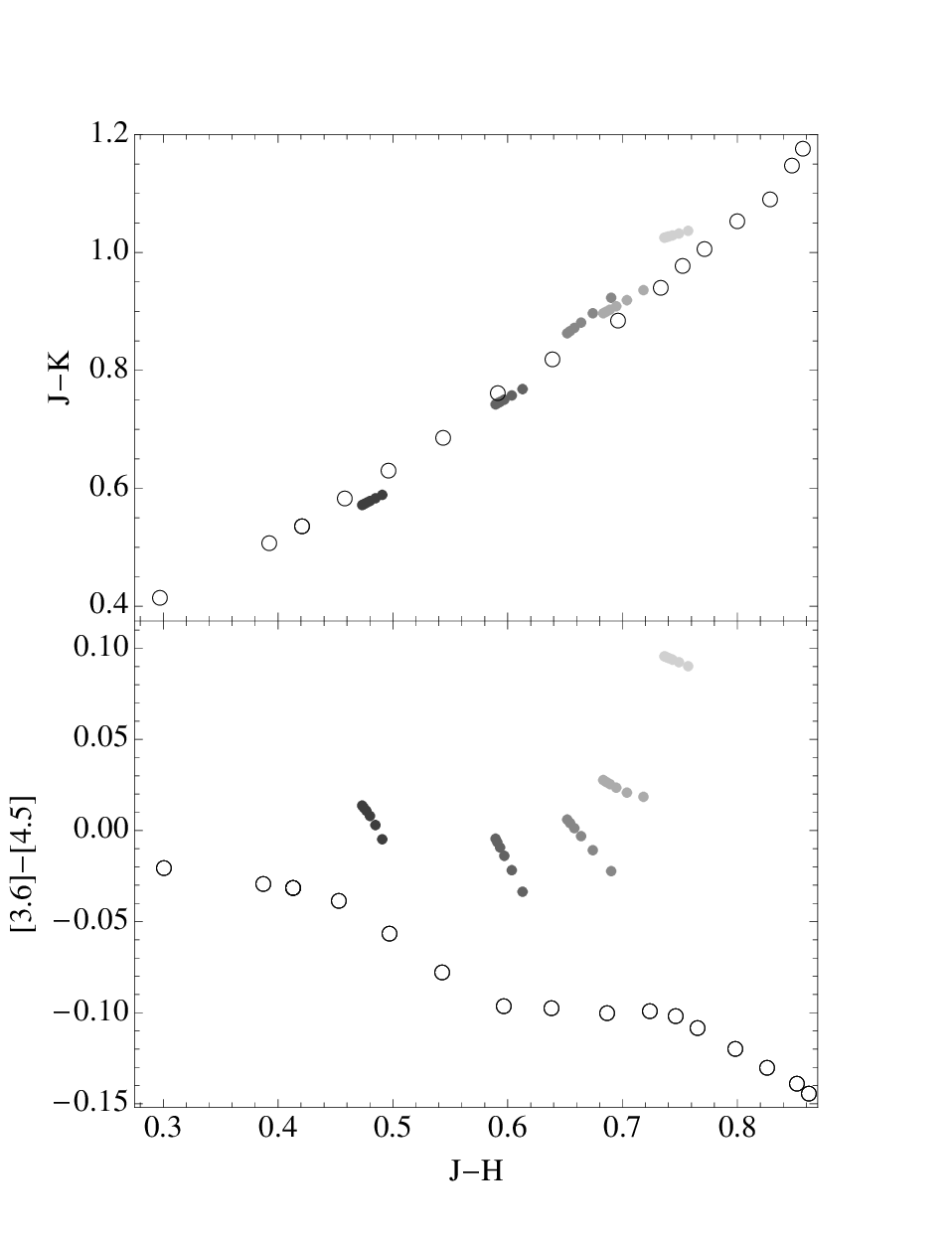}
\caption{Comparison between the NIR colors of giant stars (open circles) and BC03 SPS models (filled circles) considered in Figures \ref{fig:bothMLgrids} and \ref{fig:iracMLgrids}.  The gray scaling of the model points varies with SPS metallicity from low (black) to high (light gray).  At each metallicity, a range of ages is shown (see Figure  \ref{fig:iracMLgrids}).   }
\label{fig:modvbb88}
\end{figure}

Again, because we are sampling the Rayleigh-Jeans tail, we expect the relations {\it between} NIR giant colors to remain robust to changes in metallicity (although the individual colors themselves may vary), and this appears to be true for our empirical calibration of the [3.6]-[4.5] color for each spectral type.  This also implies that we can use these relations to cast the NIR colors of synthesized stellar population models in terms of the [3.6]-[4.5] color.  Indeed, in Figure \ref{fig:modvbb88} we show the NIR colors of the 15 BB88 spectral types together with those of synthesized stellar populations that span a range of metallicities and ages (first introduced in $\S$ \ref{sec:withcolor} in the main text).  The close agreement stems from the fact that the colors of our old synthesized stellar populations are dominated by the colors of giants, which have much higher luminosities than dwarf stars (as much as 2 orders of magnitude).  
\begin{table}[t]
\begin{center}
\caption{Best-fitting relations between the NIR and [3.6]-[4.5] colors of giant stars in GLIMPSE\label{tab:glimpsefits}}
\begin{tabular}{rcccc}
\tableline\tableline
 color&$a$&$b$&$\delta$$a$&$\delta$$b$\\
\tableline
J-H&-0.223&0.053&0.013&0.007\\
J-K&-0.165&0.049&0.01&0.008\\
H-K&-0.604&0.031&0.054&0.031\\
\tableline
\end{tabular}
\tablecomments{\small {Parameters of the best-fit relation [3.6]-[4.5]=$a$ (color)+ $b$ to 15 giant spectral classes defined in BB88.   }}
\end{center}
\end{table}


\newpage
\begin{thebibliography}{}
\bibitem[Bakes et al.(2001)]{bakes}Bakes, E. L. O., Tielens, A. G. G. M., Bauschlicher, C. W., Jr., Hudgins, D. M. \& Allamandola, L. J., 2001, ApJ, 560, 261
\bibitem[Bell \& de Jong(2001)]{bdJ01}Bell, E. F. \& de Jong, R. S.  2001, ApJ, 550, 212
\bibitem[Benjamin et al.(2003)]{benjamin}Benjamin, B. A., Churchwell, E., Babler, B. L. et al. 2003, PASP, 115, 953
\bibitem[Bernardi et al.(2010)]{bernardi}Bernardi, M., Shankar, F., Hyde, J. B., Mei, S., Marulli, F., Sheth, R. K., 2010, MNRAS, 404, 2087
\bibitem[Bessell \& Brett(1988)]{bb88}Bessell, M. S. \& Brett, J. M. 1988PASP, 100, 1134
\bibitem[Binney \& Merrifield(1998)]{binmerr}Binney, J., \& Merrifield, M. 1998, Galactic Astronomy (Princeton, NJ: Princeton University Press) (Princeton series in astrophysics) QB857.B522
\bibitem[Bruzual \& Charlot(2003)]{bc03}Bruzual G. \& Charlot S., 2003, MNRAS, 344, 1000
\bibitem[Bruzual (2007)]{cb07}Bruzual, G. ArXiv Astrophysics e-prints, 2007, astro-ph/0703052
\bibitem[Buta et al.(2010)]{buta10}Buta R., et al., 2010, ApJS, 190, 147
\bibitem[Cardelli et al.(1989)]{cardelli}Cardelli, J. A., Clayton, G. C. \& Mathis, J. S. 2001, 1989, ApJ, 345, 245
\bibitem[Carpenter(2001)]{carpenter}Carpenter, J. M., 2001, AJ, 121, 2851
\bibitem[Cesetti et al.(2013)]{cesetti}Cesetti, M., Pizzella, A., Ivanov, V. D. et al. 2013, A\&A, 549, 129
\bibitem[Churchwell et al.(2009)]{churchwell}Churchwell, E., Babler, B. L., Meade, M. R. et al. 2009, PASP, 121, 213
\bibitem[Courteau et al.(1996)]{courteau}Courteau, S., de Jong, R. S., \& Broeils, A. H. 1996, ApJL, 457, 73
\bibitem[da Cunha et al.(2008)]{dacunha}da Cunha E., Charlot S. \& Elbaz D., 2008, MNRAS, 388, 1595
\bibitem[Davidge(2009)]{davidge}Davidge, T. J. 2009, ApJ, 697, 1439
\bibitem[Draine(2003)]{draine}Draine, B. T., 2003, Annu. Rev. Astron. Astrophys., 41, 241
\bibitem[de Blok et al.(2008)]{deblok}de Blok, W.J.G., Walter, F., Brinks, E., Trachternach, C., Oh, S-H. \& Kennicutt, R.C. 2008, AJ, 136, 2648
\bibitem[de Denus-Baillargeon et al.(2013)]{dedenus}de Denus-Baillargeon, M.-M., Hernandez, O., Boissier, et al. 2013, astro-ph/1307.2516
\bibitem[Elmegreen \& Elmegreen(1984)]{ee84}Elmegreen, D. M., Elmegreen, B. G., 1984, ApJS, 54, 127
\bibitem[Elmegreen et al.(2007)]{elm07}Elmegreen, B. G., Elmegreen, D. M., Knapen, J. H., Buta, R. J., Block, D. L. \& Puerari, I. 2007, ApJL, 670, 97
\bibitem[Eskew, Zaritsky \& Meidt(2012)]{eskew}Eskew, M., Zaritsky, D. \& Meidt, S. E. 2012, AJ, 143 139
\bibitem[Ferreras et al.(1999)]{ferreras}Ferreras, I., Charlot, S. \& Silk, J. 1999, ApJ, 521, 81
\bibitem[Flagey et al.(2006)]{flagey}Flagey N., Boulanger F., Verstraete L., Miville Deschnes M. A., Noriega Crespo A., Reach W. T., 2006, A\&A, 453, 969
\bibitem[Florido et al.(2001)]{florido}Florido, E., Battaner, E., Guijarro, A., Garz\'on, F., \& Jim\'enez-Vicente, J. 2001, A\&A, 378, 82
\bibitem[Foyle et al.(2010)]{foyle}Foyle, K., Rix, H.-W. \& Zibetti, Z. 2010, MNRAS, 407, 163
\bibitem[Gallazi et al.(2005)]{gallazi1}Gallazzi, A., Charlot, S., Brinchmann, J., White, S. D. M. \& Tremonti, C. A. 2005, MNRAS, 362, 41
\bibitem[Gallazzi et al.(2006)]{gallazi}Gallazzi, A., Charlot, S., Brinchmann, J., \& White, S. D. M., 2006, MNRAS, 370, 1106
\bibitem[Ganda et al.(2009)]{ganda}Ganda, K., Peletier, R. F., Balcells, M. \& Falcon-Barroso, J. 2009, MNRAS, 395, 1669
\bibitem[Governato et al.(2007)]{governato}Governato, F., Willman, B., Mayer, L., Brooks, A., Stinson, G., Valenzuela, O., Wadsley, J., \& Quinn, T. 2007, MNRAS, 374, 1479
\bibitem[Guo \& White(2008)]{guo}Guo, Q. \& White, S. D. 2008, MNRAS, 384, 2
\bibitem[Haan(]{haan}Haan, S. Schinnerer, E., Emsellem, E., Garcia-Burillo, S., Combes, F., 
Mundell, C. G. \& Rix, H.-W. 2009, ApJ, 692, 1623
\bibitem[Harris \& Zaritsky(2009)]{HZ}Harris, J. \& Zaritsky, D. 2009, AJ, 138, 1243
\bibitem[Hauschildt et al.(1999)]{hauschildt}Hauschildt, P. H., Allard, F., Ferguson, J., Baron, E., \& Alexander, D. R. 1999, ApJ 525, 195
\bibitem[Hunter et al.(2006)]{hunter}Hunter, D. A., Elmegreen, B. G. \& Martin, E. 2006, AJ, 132, 801
\bibitem[Hyv\"arinen(1999)]{hica}Hyv\"arinen A., 1999, IEEE Signal Processing Lett. , 6, 145
\bibitem[Hyv\"arinen \& Oja(2000)]{hoica}Hyv\"arinen, A. \& Oja, E., 2000, Neural Networks , 13, 411 
\bibitem[Indebetouw et al.(2005)]{indebetouw}Indebetouw, R., Mathis, J. S., Babler, B. L., et al. 2005, ApJ, 619, 931 
\bibitem[Jansen et al.(1994)]{jansen}Jansen, R. A., Knapen, J. H., Beckman, J. E., Peletier, R. F., \& Hes, R. 1994, MNRAS, 270, 373
\bibitem[Jarrett et al.(2000)]{jarrett}Jarrett T. H., Chester T., Cutri R., Schneider S., Skrutskie M., Huchra J. P., 2000, AJ, 119, 2498
\bibitem[Jarrett et al.(2013)]{jarrett2013}Jarrett T. H., Masci, F., Tsai, C. W. et al. 2013, AJ, 145, 1538
\bibitem[MacArthur et al.(2004)]{macarthur}Macarthur, L. A., Courteau, S., Bell, E. \& Holtzman, J. A. 2004, ApJSS, 152,175
\bibitem[Marigo et al.(2008)]{marigo}Marigo P., Girardi L., Bressan A., Groenewegen M. A. T., Silva L., Granato G. L., 2008, A\&A, 482, 883
\bibitem[Meidt et al.(2012a)]{meidtpaperI}Meidt S. E., Schinnerer, E., Knapen, J., et al. 2012, ApJ, 744, 17
\bibitem[Meidt et al.(2012b)]{meidt2012b}Meidt S. E., Schinnerer, E., Munoz-Mateos, J. C., et al. 2012, ApJ, 748, 30
\bibitem[Mouhcine \& Lan\c{c}on(2002)]{lancon}Mouhcine, M.  \& Lan\c{c}on, A. 2002, A\&A, 393, 149
\bibitem[Leroy et al.(2008)]{leroy}Leroy, A., Walter, F., Brinks, E., Bigiel, F., de Blok, W. J. G., Madore, B. \& Thornley, M. D. 2008, AJ, 136, 2782
\bibitem[Kendall et al.(2008)]{kendall}Kendall, S., Kennicutt, R. C., Clarke, C. \& Thornley, M. D. 2008, MNRAS, 387, 1007
\bibitem[Kennicutt et al.(2003)]{kennSings}Kennicutt, Jr., R. C., et al. 2003, PASP, 115, 928
\bibitem[Kenyon \& Hartmann(1995)]{kenyon}Kenyon, S. J. \& Hartmann, L. 1995, ApJS, 101, 117
\bibitem[Kreckel et al.(2013)]{kreckel}Kreckel, K. Groves, B., Schinnerer, E. et al. 2013, ApJ, 771, 62
\bibitem[Kroupa(2001)]{kroupa}Kroupa, P. 2001, MNRAS, 322, 231
\bibitem[Maraston(2005)]{maraston05}Maraston, C. 2005, MNRAS, 362, 799
\bibitem[Mouri \& Taniguchi(2000)]{mouri}Mouri, H. \& Taniguchi, Y., 2000, ApJL, 545, 103
\bibitem[Navarro \& White(1994)]{navarro}Navarro, J. F. \& White, S. D. M. 1994, MNRAS, 267, 401
\bibitem[Oh et al.(2008)]{oh}Oh, S.-H., de Blok, W. J. G., Walter, F., Brinks, E. \& Kennicutt, R. C. 2008, AJ, 136, 2761
\bibitem[Origlia et al.(2010)]{origlia}Origlia, O., Rood, R. T.,  Fabbri, S., et al. 2010, ApJ, 718, 522
\bibitem[Pahre et al.(2004a)]{pahre2}Pahre, M. A., Ashby, M. L. N., Fazio, G. G., \& Willner, S. P. 2004a, ApJS, 154, 229
\bibitem[Pahre et al.(2004b)]{pahre}Pahre, M. A., Ashby, M. L. N., Fazio, G. G., \& Willner, S. P. 2004b, ApJS, 154, 235
\bibitem[Peletier et al.(2012)]{peletier}Peletier, R., F., Kutdemir, E., van der Wolk, G. et al. 2012, MNRAS, 419, 2031
\bibitem[Popescu et al.(2011)]{popescu}Popescu, C. C., Tuffs, R. J., Dopita, M. A., et al. 2011, A\&A, 527, 109
\bibitem[Prescott et al.(2007)]{prescott}Prescott, M. K. M., Kenicutt, R. C., Bendo, G. J. et al. 2007, ApJ, 668, 182
\bibitem[Reach et al.(2005)]{reach}Reach, W. T., et al. 2005, PASP, 117, 978
\bibitem[Querejeta et al.(2014)]{querejeta}Querejeta, M., Meidt, S. E., Schinnerer, E., et al. 2014, in prep. 
\bibitem[Salo(2010)]{salo}Salo, H. 2010, http://sun3.oulu.fi/~hsalo/galfidl.html
\bibitem[Sheth et al.(2010)]{sheth}Sheth, K. et al. 2010, submitted to PASP
\bibitem[Somerville(2002)]{somerville}Somerville, R. S. 2002, ApJL, 572, 23
\bibitem[Temi et al.(2008)]{temi}Temi, P., Brighenti, F. \& Mathews, W. G. 2008, ApJ, 672, 244
\bibitem[Willner et al.(2004)]{willner}Willner, S. P. et al. 2004, ApJSS, 154,222
\bibitem[Zhang \& Buta(2007)]{zhang}Zhang, X. \& Buta, R. 2007, AJ, 133, 2584
\bibitem[Zhu et al.(2010)]{zhu}Zhu, Y. N., Wu, H., Li, H.-N. \& Cao, C. 2010, A\&A, 329, 1674
\bibitem[Zibetti et al.(2009)]{z09}Zibetti, S., Charlot, S. \& Rix, H. -W. 2009, MNRAS, 400, 1181
\end{thebibliography}
\end{document}